\documentclass[prl,aps,twocolumn,showpacs]{revtex4} 
\usepackage{graphicx} 
\usepackage{tabularx}
\usepackage{dcolumn} 
\usepackage{color}
\usepackage{bm}
\usepackage{amsmath}
\usepackage{amssymb}

\begin{document} 
 
\title{%
Spin triplet nematic pairing symmetry and superconducting double transition in U$_{1-x}$Th$_{x}$Be$_{13}$
}

\author{Kazushige Machida} 
\affiliation{Department of Physics, Ritsumeikan University, 
Kusatsu 525-8577, Japan} 

\date{\today}

\begin{abstract}
Motivated by a recent experiment on U$_{1-x}$Th$_{x}$Be$_{13}$ with $x=3\%$,
we develop a theory to narrow down the possible pair symmetry to consistently describe the double transition
utilizing various theoretical tools; group theory and Ginzburg-Landau theory.
It is explained in terms of the two dimensional representation E$_{\rm u}$ with spin triplet.
A symmetry breaking causes the degenerate $T_{\rm c}$ to split into the two.
The low temperature phase is identified as the cyclic $p$ wave: $\vec {d}({\bf k})={\hat x}k_x+\varepsilon{\hat y}k_y+\varepsilon^2{\hat z}k_z$ 
with $\varepsilon^3=1$ while the biaxial nematic phase: ${\vec d}({\bf k})={\sqrt 3}({\hat x}k_x-{\hat y}k_y$) is the high temperature one.
This allows us to simultaneously identify 
the uniaxial nematic phase: ${\vec d}({\bf k})=2{\hat z}k_z-{\hat x}k_x-{\hat y}k_y$ for UBe$_{13}$, which breaks spontaneously cubic symmetry of the system.
Those pair functions are fully consistent with the above and existing data.
We comment on the accidental scenario in addition to this degeneracy scenario and the intriguing topological nature hidden in this long-known material.
\end{abstract}

 
 
\maketitle 



It is generally quite difficult to uniquely identify the pairing symmetry among 
abundant possible states in exotic superconductors and superfluids~\cite{review1,review2}.
This is particularly true for the spin-triplet pairing, compared with the spin-singlet one because
the lenlarged degrees of freedom in a pair function consisting of the spin and orbital components
are doubled to the singlet case where the former freedom is absent.
Among many unconventional heavy fermion superconductors whose pairing symmetry is largely
undetermined precisely, UPt$_3$ is a rare case to identify a triplet pairing function.
This is partly because of the double transition at $H=0$ and the multiple phase in the $H$ vs $T$
plane. In order to explain the multiple phase in UPt$_3$ it is obvious that
the pairing function must belong to a multi-dimensional representation,
namely a two-dimensional one E$_{1{\rm u}}$~\cite{tsutsumi,machida} or E$_{2{\rm u}}$~\cite{sauls0,nishira}
in D$_{6h}$ whose orbital degeneracy can be lifted by an 
external or internal symmetry  breaking perturbation, that is, the antiferromagnetism in this case,
giving rise to the double transition at $T_{{\rm c}1}$ and $T_{{\rm c}2}$.

There is the other well-known double transition superconductor (U$_{1-x}$Th$_{x}$)Be$_{13}$
for a narrow doping region: $2\%<x<4\%$ (see Fig. 1) whose pairing symmetry has not been identified yet~\cite{sigrist-ueda} since its
discovery in 1985~\cite{ott}. This long-standing mystery remains unexplored because the difficulty of high quality
single crystal synthesis prevents its detailed investigations. Almost all experiments on this system
are done by using poly-crystalline materials except for a few example~\cite{batlogg,heffner,sonier}.

Recently a remarkable experiment~\cite{shimizu1} has been reported by using single crystalline (U$_{0.97}$Th$_{0.03}$)Be$_{13}$
with optimal doping. The unprecedented field angle resolved specific heat experiment reveals the following new facts
which prompt us to renew the view accumulated so far:
 
 \noindent
 (1) The gap structure is nodeless for the low temperature phase which is the same as in the parent compound
 UBe$_{13}$ reported earlier~\cite{shimizu}. 

 \noindent
 (2) Under applied fields, the two transition lines starting at the two transition temperatures, $T_{{\rm c}1}$=0.6K and $T_{{\rm c}2}$=0.4K
 do not exhibit a crossing (see the inset of Fig. 1) as coinciding with the poly-crystalline experiment with the comparable doping level by 
 Kromer et al~\cite{kromer}. We call A (B) phase for the high (low) temperature phase in the $H$ vs $T$ plane.
 This is sharply contrasted with UPt$_3$ where in the $H$ vs $T$ plane the phase diagram consists of the three 
 subphases. This contrast results in important consequences in choosing the possible pair function
 because as we will see shortly, the orbital degenerate state in cubic symmetry generically removes the crossing
 in a fundamental way deeply rooted into the group theoretical structure for O$_h$, differing from D$_{6h}$
 in UPt$_3$~\cite{tsutsumi,machida,sauls,nishira}. Thus the two phases A and B in the $H$ vs $T$ plane are inevitable in O$_h$.

\noindent
 (3) The $H$ vs $T$ phase diagram is almost independent of the field orientation, thus enabling us to trust the phase diagram
 constructed by poly-crystalline samples \cite{kromer}.

\noindent
 (4) The field angle resolved specific heat experiment exhibits a different orientational dependence for 
 A and B phases, unambiguously indicating that A and B phase are different superconducting states
 with its own gap structures assigned for each phase.

Let us now start out by examining the possible pairing state to explain the facts (1)-(4) consistently.
In order to naturally explain the two phases A and B, we need a pair function
 belonging to a higher dimensional representation, at least two dimensional one,
 which is the minimal dimension. We call it the degenerate scenario.
 If we choose the pair function from the one dimensional representation, say for A phase,
 we must select the other for B phase belonging to a different irreducible representation.
 We call it the accidental scenario, which is extensively studied earlier~\cite{sigrist}.

\begin{figure}
\includegraphics[width=6cm]{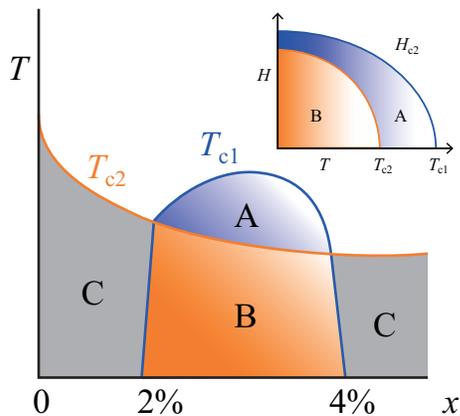}
\caption{(color online)
Schematic phase diagram in $T$-$x$ plane of U$_{1-x}$Th$_{x}$Be$_{13}$. 
A: the biaxial nematic state $\vec {d}({\bf k})=\sqrt{3}({\hat x}k_x-{\hat y}k_y)$,
B: the cyclic $p$ state $\vec {d}({\bf k})={\hat x}k_x+\varepsilon{\hat y}k_y+\varepsilon^2{\hat z}k_z$,
C: the uniaxial nematic state $\vec {d}({\bf k})=2{\hat z}k_z-{\hat x}k_x-{\hat y}k_y$  for E$_{\rm u}$.
The inset shows schamatic phase diagram for $2\%<x<4\%$ in $H$-$T$ plane. The A-B boundary never touchs the H$_{c2}$ line
generically.
}
\end{figure}

 As mentioned, B phase must be nodeless. For A phase it is a different gap structure 
 from this because they responds differently to external fields.
According to $\mu$SR experiment~\cite{heffner0}, B phase is a time reversal broken state
 while A phase is an unbroken state. Previously the origin of this spontaneous appearance 
 of the internal magnetic field was assigned to a spin density wave state~\cite{heffner0,kato1,kato2,kato3}.
 The later extensive search in the reciprocal space of the magnetic Bragg peaks by
 neutron experiments~\cite{magne} fails to find it. This proposal is no longer applicable to 
 the present system.
Thus we must consider the double transition as arising the successive phase transition between 
different superconducting states. In retrospect, it is obvious that the second transition at $T_{\rm c2}$
is a superconducting transition because $H_{{\rm c1}}$ is enhanced below $T_{\rm c2}$~\cite{heffner0},
meaning that the superconducting condensation energy is gained at lower temperatures
by going into a more stable superconducting state.

Another crucial $\mu$SR experiments~\cite{heffner,sonier} done by single crystals
 on undoped and 3\% Th-doped systems show a clear decrease of the Knight shift below the
 superconducting transition. In both cases $\chi_{\rm s}/\chi_{\rm N}$ tends to $\sim 0.5$
 with $\chi_{\rm s}$ ($\chi_{\rm N}$) spin susceptivity in the superconducting (normal) state
 in common to lower temperatures.  In the doped system $\chi_{\rm s}$ exhibits
 a clear anomaly at $T_{\rm c2}$, signaling that A and B phases have a different spin structure.
 Thus A and B phases possess not only different orbital structures, but also different 
 spin structures in the pair symmetry.

 This result can be interpreted by the two ways:
 The triplet pair case can explain this decrease of the spin susceptibility $\chi_{\rm s}\rightarrow 0.5\chi_{\rm N}$
towards $T\rightarrow0$ if the $d$-vector consists of three components, such as in superfluid $^3$He B phase
where the pair function is described by ${\vec d}({\bf k})={\hat x}k_x+{\hat y}k_y+{\hat z}k_z$ 
with $\chi_{\rm s}\rightarrow {2\over3}\chi_{\rm N}$ at $T=0$. Note that $\chi_{\rm s}\rightarrow 0.3\chi_{\rm N}$
in the actual B phase because of the Fermi liquid correction~\cite{leggett}.
Thus our B phase should have multiple $d$-vector components under the single domain assumption, which we assume
in the following arguments.


\begin{center}
\begin{table}[h]
\begin{tabular}{cccc} 
\multicolumn{4}{c}{Characteization of the phases}\\  \hline
phase & d-vector & gap & $\chi_s/\chi_N$ \\  \hline
A (BN)&$\sqrt{3}({\hat x}k_x-{\hat y}k_y)$& point nodes & ${1/2}$($H \perp z$) 1($\parallel z$)\\ 
B (Cyclic)&${\hat x}k_x+\varepsilon{\hat y}k_y+\varepsilon^2{\hat z}k_z$& points \& full &${2/3}$\\ 
C (UN) &$2{\hat z}k_z-{\hat x}k_x-{\hat y}k_y$&full gap &${2/3}$\\  \hline
\end{tabular}
\caption{\label{table} BN: biaxial nematic, UN: uniaxial nematic, Cyclic: p wave cyclic}
\end{table}
\end{center}

 The singlet pair case might also explain this because $\chi_{\rm s}/\chi_{\rm N}\rightarrow$ a finite value 
 by assuming the impurity effect, which gives rise to a finite residual density of states.
 However, it is rather strange to see the similar values $\chi_{\rm s}\rightarrow 0.5\chi_{\rm N}$
 for the completely different systems, undoped and 3\% Th-doped systems.
 We support the former interpretation, namely the triplet pair case which is fully backed up by
 the arguments followed.

In the $x$-$T$ phase diagram of (U$_{1-x}$Th$_{x}$)Be$_{13}$ shown in Fig.~1,
 we can assign each phase to A, B, and C. Here we assume that the phase boundaries are all second order.
 Under pressure $P$ for $x=2.2\%$ Zieve et al~\cite{zieve} found that the four lines meet at a point in the $P$-$T$ plane.
 According to the 2D degenerate scenario which we advocate here, the two components of the 2D representation
 corresponding to A and C, and its linear combined state to B phase.
 Then it is logical to consider that the high doping phase above $x>4\%$ must be C phase,
 exhausting all possible combinations allowed by the 2D degenerate scenario.
 That is, the undoped UBe$_{13}$ and higher Th dopings must be the identical phase C.
 This is a ``minimal'' phase diagram. Note a piece of supporting evidence
 that the undoped and 6\% Th doped systems are a time reversal symmetry unbroken state,
 different from B, but same as A phase~\cite{heffner0}.
 This 2D degenerate scenario fundamentally differs from the accidental scenario,
 where generically the higher doping phase is not necessarily the same as in the undoped 
 UBe$_{13}$ from the logical point of view.
 That is, it could be a D phase in the accidental scenario because there is no internal constraint here.
 
 Let us now to go on along this line to further narrow down possible pairing symmetry.
 The group theoretic classification under cubic symmetry O$_{\rm h}$,
 which has been done by several groups~\cite{ozaki,gorkov,ueda,sigrist-ueda}.
 Among the 2D representations E$_{\rm g}$ for singlet state and E$_{\rm u}$ for triplet state,
 we first examine the triplet case; $\vec {l}_1({\bf k})=\sqrt{3}({\hat x}k_x-{\hat y}k_y)$,
$\vec {l}_2({\bf k})=2{\hat z}k_z-{\hat x}k_x-{\hat y}k_y$
with the standard notations~\cite{sigrist-ueda}
where $l_1(l_2)$ is known to be the biaxial (uniaxial) nematic state.
Those break the cubic symmetry down to $C_2$.
A combination of $\vec {l}_1({\bf k})+i\vec {l}_2({\bf k})$ is casted into
$\vec {d}({\bf k})={\hat x}k_x+\varepsilon{\hat y}k_y+\varepsilon^2{\hat z}k_z$ 
with $\varepsilon^3=1$. We call it the cyclic $p$ wave  state, 
which is a so-called inert state, namely generically stable state against parameter's change of a system~\cite{ozaki}.

Analogously, we can find the corresponding singlet state, namely
${l}^s_1({\bf k})=\sqrt{3}(k_x^2-k_y^2)$, and 
${l}^s_2({\bf k})=2k_z^2-k_x^2-k_y^2$.
with a linear combination ${l}^s_1({\bf k})+i{l}^s_2({\bf k})$, giving rise to
$\psi({\bf k})=k_x^2+\varepsilon k_y^2+\varepsilon^2 k_z^2$. 
This is the cyclic $d$ wave state.
Here we examine those comparatively.

The gap structure of the cyclic $p$ wave, which is a non-unitary state, is easily calculated: One branch has
 point nodes at (111) direction and its equivalent directions, and the other branch is a full gap.
 Since the Fermi surface is absent along the (111) direction according to band calculations~\cite{take,maehira},
 this state is actually no node on the Fermi surface.
 This is fully consistent with Shimizu's experiment~\cite{shimizu1}.
 The cyclic $d$ wave state has point nodes at (111) direction, just the same as the above,
 it is also a candidate for B phase. However, ${l}^s_1({\bf k})$ and ${l}^s_2({\bf k})$ have line nodes,
 one of which is going to identify C phase, thus it is not a candidate pairing state because UBe$_{13}$
 is a full gap. Thus E$_{\rm g}$ degenerate scenario does not hold anymore.
 We are left with the E$_{\rm u}$ degenerate scenario, which we consider from now on.
 
 The Ginzburg-Landau free energy  density $F$ for the 2D representation is written down
$F=\alpha(T)(|\vec {l}_1|^2+|\vec {l}_2|^2)+
\beta_1(|\vec {l}_1|^2+|\vec {l}_2|^2)^2+\beta_2(\vec {l}_1\vec {l}^{\ast}_2-\vec {l}^{\ast}_1\vec {l}_2)^2$
with $\alpha(T)=\alpha_0(T_{\rm c}-T)$.
The degenerate transition temperature $T_{\rm c}$ can be split into two $T_{{\rm c}1}$ and $T_{{\rm c}2}$
whose origin is unknown at present.  We will discuss a possible origin later.
The condition of $\beta_1>0$ is required for the stability 
and if $\beta_2<0$, it is easy to see that $l_1({\bf k})+il_2({\bf k})$, or the cyclic $p$ wave state is stable in lower temperatures as a ground state.
Thus this time reversal broken state is realized. The weak coupling estimate
for $p$ wave case~\cite{ueda} shows that $\beta_1: \beta_2=1: (-1/3)$.

Thus $\vec {l}_1({\bf k})$ first appears at $T_{{\rm c}1}$ as A phase (the biaxial nematic)
and $\vec {l}_1({\bf k})+i\vec {l}_2({\bf k})$ with $\chi_s/\chi_N$=2/3 for all field directions
appear at a lower $T_{{\rm c}2}$ as B phase (the cyclic $p$ wave) via a second order transition.
Logically in the $x$-$T$ plane the remaining C phase (uniaxial nematic) must be $\vec {l}_2({\bf k})$.
This identification is quite appealing because  $\vec {l}_2({\bf k})=2{\hat z}k_z-{\hat x}k_x-{\hat y}k_y$
is full gap independent of the Fermi surface topology and $\chi_s/\chi_N$=2/3 for all field directions.
For A phase $\vec {l}_1({\bf k})=\sqrt{3}({\hat x}k_x-{\hat y}k_y)$ has two point nodes at
the poles with $\chi_s/\chi_N$=1/2 for $H\parallel x$ and $y$. 
Those spin structures of A, B and C phases are fully consistent with the $\mu$SR results~\cite{heffner,sonier}.
So far there is no experimental information on the precise gap structure for A phase.

Several researchers have explored and been characterizing the nature of the identified triplet states in a different context:
All those states belong to the so-called ${^3}P_2$ superfluid phase ($S=1, L=1, J=2$)
which is thought to be realized in the neutron-rich inner core of the neutron stars~\cite{tamagaki,mizushima1}.
Those have all topologically rich structure since the two nematic states are classified to DIII in the topological table~\cite{furusaki}
and are similar to superfluid $^3$He B phase where the surface Majorana mode exists~\cite{review2}.
The cyclic $p$ state has the Weyl nodes in general whose surface Majorana arc structure is
discussed~\cite{mizushima2}.
We also point out that the recent so-called ``nematic superconductors'', which spontaneously break the rotational symmetry down to C$_2$,
such as A$_x$Bi$_2$Se$_3$ (A=Cu, Sr, Nb)~\cite{matano,yonezawa,visser} are discussed in terms of the $p$ wave pairing state~\cite{fu}
similar to the present nematic states, sharing the same physics.

Now we consider the $H$-$T$ phase diagram. It can be discussed by the following 
GL free energy density $F$ for 2D representation in O$_{\rm h}$~\cite{machida0}:


\begin{eqnarray}
&F\!=\! K_1\{|D_x(\!-\!l_1\!+\!{\sqrt3}l_2)|^2\!+\!|D_y(l_1\!+\!{\sqrt3}l_2)|^2\!+\!|D_z(2l_1)|^2\} 
\nonumber \\
&\!+\!K_2\{|D_x({\sqrt3}l_1\!+\!l_2)|^2\!+\!|D_y(-\!{\sqrt3}l_1\!+\!l_2)|^2\!+\!|D_z(2l_1)|^2\}
\nonumber \\
&+\Sigma{_{i=1,2}} \alpha_{i}(T)|l_i|^2
\end{eqnarray}


\noindent
with $\alpha_{i}(T)=\alpha_0 (T-T_{ci})$. The covariant derivative $\vec{D}=(\partial_x, \partial_y,-2iA_z (x,y))$
with $\hbar=c=e=1$ and $A_z(x,y)=H(-x\sin\theta+y\cos\theta)$ 
for the magnetic field $H$ applied in the $x$-$y$ plane and $\theta$ is the angle
 from the $x$-axis. Then, the total free energy ${\mathcal F}$ is given

\begin{eqnarray}
{\mathcal F}=\int^{\infty}_{-\infty}du\{(K_1+3K_2)({dl_1\over du})^2+4K_1u^2l_1^2
\nonumber \\
+(3K_1+K_2)({dl_2\over du})^2+4K_1u^2l_2^2
\nonumber \\
-2\sqrt3(K_1-K_2)\cos2\theta({dl_1\over du})({dl_2\over du})+\Sigma{_{i=1,2}} \alpha_{i}(T)|l_i|^2\}
\end{eqnarray} 

\noindent
From this form it is clear that because of the bilinear coupling between $l_1$
and $l_2$ the two transition lines $H_{\rm c2}^{(1)}$ and $H_{\rm c2}^{(2)}$
starting at $T_{{\rm c}1}$ and $T_{{\rm c}2}$ respectively never cross generically.
It is interesting to note that $K_1=K_2$ is rather a special case because the gradient coupling 
term above vanishes and the initial slopes of the two $H_{\rm c2}$ curves become same, so they
are exactly parallel in the $H$-$T$ plane. Thus within this scenario the two $H_{\rm c2}$ curves
do not cross to each other in general. This is the case where according to Kromer et al~\cite{kromer}
all the experimental phase diagrams for various dopings show no crossing.
They are more or less parallel shifted~\cite{rosenbaum}.

When $K_1\neq K_2$, the second transition at $T_{{\rm c}2}$ becomes a crossover
differed from the true second order at $H=0$ simply because below $T_{{\rm c}1}$ the
second component $l_2$ is always induced by $l_1$ via the bilinear coupling above.
Therefore, to keep the second order transition for $H\neq 0$ the condition $K_1\simeq K_2$
is desirable.
It can be derived~\cite{machida0} when $T_{{\rm c}1}$=$T_{{\rm c}2}$ the following in-plane $H_{\rm c2}(\theta)$
anisotropy

\begin{equation}
{H_{\rm c2}(\theta)\over H_{\rm c2}(0)}=\left[{{2(K_1+K_2)+|K_1-K_2|}\over{2(K_1+K_2)+|K_1-K_2|\sqrt{1+3\cos^22\theta}}}\right]^{1/2}.
\end{equation}


\noindent
When $K_1= K_2$, this in-plane anisotropy vanishes.
The two $H_{\rm c2}$ curves are parallel in the $H$-$T$ plane
and completely no in-plane anisotropy, both of which are second order.

The expression for $K_1/K_2$ is given in terms of the Fermi velocity average over the
Fermi surface, $K_1/K_2=\left[4\langle v_z^4\rangle+2\langle v_z^2\rangle\langle v_x^2\rangle\right]/6\langle v_z^2 v_x^2\rangle$.
For the Fermi sphere model, $\langle v_z^2\rangle=1/5$ and $\langle v_z^2v_x^2\rangle=1/15$,
thus $K_1/K_2=13/6=2.2$ far from unity. However, if we approximate the octahedron Fermi surface with holes
along (111) direction at the $\Gamma$ point~\cite{take,maehira}
to a complete regular octahedron, we obtain $K_1/K_2=1$.
The continuous deformation of this octahedron might change $K_1/K_2=1$
towards $K_1/K_2=2.2$ for the sphere. Thus it is not unreasonable to expect 
$K_1/K_2\simeq 1$, which assures the second order like transition at $T_{\rm c2}$
for $H\neq 0$. Experimentally there is no indication of the first order transition or crossover
phenomenon at $H_{\rm c2}$~\cite{shimizu1}.

Let us briefly compare the UPt$_3$ problem in hexagonal D$_{6\rm h}$ with the present 
cubic O$_{\rm h}$.
In the former the gradient terms consist of the four independent invariants~\cite{machida} with the coefficients
$K_1$, $K_2$, $K_3$, and $K_4$ where the latter two terms cause the gradient coupling. 
To explain the observed crossing, ultimately giving rise to the three phases in the $H$-$T$ plane of UPt$_3$, 
$K_3$ and $K_4$ need a fine tuning to be $K_3$=$K_4$=0.
In UPt$_3$  the spin degeneracy scenario seems to be more natural~\cite{machida2,machida3} than the orbital one.
In O$_{\rm h}$ case here, since there are only two independent parameters $K_1$ and $K_2$,
the gradient coupling is inevitable except for the special point $K_1=K_2$.
Therefore the two $H_{\rm c2}$ curves intrinsically avoid each other without any fine tuning.


So far we focus on the 2D representation E$_{\rm u}$ and E$_{\rm g}$.
There are two three dimensional representations both for triplet and singlet cases~\cite{sigrist-ueda}.
But it is clear that no appropriate state is found in light of our criteria:
nodal structure and time reversal broken state for B phase.
Moreover, there are many other redundant states more than the observed A, 
B and C phases by combing the three components.
Thus we do not pursue this possibility, concluding that the present E$_{\rm u}$ is
the best choice.

Let us move to the accidental scenario which has been discussed extensively by Sigrist and Rice~\cite{sigrist}.
Since as mentioned we obtain new information as for the nodal structure of undoped and doped systems, 
we can advance their argument and narrow down the possible pair symmetry out of the huge numbers of the possible combinations.
As for the singlet case, we start with the simplest A$_{1\rm g}$, namely the $s$ wave state at $x=0$
as C phase, which is nodeless. Upon doping another singlet state such as A$_{2\rm g}$ appears whose $T_{\rm c 1}(x)$
curve  has a dome shape as seen from Fig.~1.
Thus at $x=3\%$ where $T_{\rm c1}(x)$ exhibits a local maximum the A$_{2\rm g}$ appears at $T_{\rm c 1}$,
then below $T_{\rm c2}$ A$_{1\rm g}$+$i$A$_{2\rm g}$ is realized as the broken time reversal state B.
This scenario seems appealing  and looks natural among vast numbers of possible combinations,
but the classified A$_{2\rm g}$ is given by $\psi(k)=(k_x^2-k_y^2)(k_y^2-k_z^2)(k_z^2-k_x^2)$, too high a angular momentum state~\cite{sigrist-ueda},
which is unlikely.
A simplest alternative is to select a $d$ wave state, instead of A$_{2\rm g}$ which is quite ad hoc,
then $s+id$ could be B phase. Note that all phases  are full gap, completely independent of the Fermi surface topology,
which is another appealing point. But there is no logical necessity for $d$-wave. 

The corresponding triplet case seems more plausible:
Starting with the simplest and most symmetric A$_{1\rm u}$, namely ${\vec d}({\bf k})={\hat x}k_x+{\hat y}k_y+{\hat z}k_z$
with full gap as C phase at $x=0$. As increasing $x$, A$_{2\rm u}$ 
${\vec d}({\bf k})={\hat x}k_x(k_y^2-k_z^2)+{\hat y}k_y(k_z^2-k_x^2)+{\hat z}k_z(k_x^2-k_y^2)$ appears to form A$_{1\rm u}$+$i$A$_{2\rm u}$
as B phase. Again those states are all full gap independent of the Fermi surface topology.
The argument is appealing because at $x=0$ the most symmetric  A$_{1\rm u}$ is realized. 
Then the doping induces the higher angular momentum A$_{2\rm u}$ by breaking some symmetry at $x=0$, such as lattice distortion
etc.

In those accidental scenarios  there is no gradient coupling by construction.
Depending on the GL parameters, the two transition lines starting at  $T_{\rm c1}$ and $T_{\rm c2}$ could be crossed.
There is no reason to avoid this crossing. Thus the $H$-$T$ phase diagram inevitably consists of three subphases,
which is not observed. Another demerit of this scenario is arbitrariness of the choice of the two possible
 independent irreducible representations A$_{1\rm u}$ and A$_{2\rm u}$. 
 
 As for the origin of the double transition, we assume some unidentified symmetry lowing perturbation
 to split $T_{\rm c}$ into $T_{\rm c1}$ and $T_{\rm c2}$.
One possible origin is crystal distortion caused by Th doping. If we consider that the distortion monotonically increases
with doping, we can not explain the non-monotonic variation of $T_{\rm c2}(x)$.
A fact~\cite{kromer} that the resistivity maximum temperature $T_{\rm max}(x)$ 
just above the transition temperature decreases with Th doping,
pointing to the $T_{\rm c2}(x)$ maximum at $x_{\rm cr}=3\%$ is quite suggestive
because the normal state properties exhibit a critical behavior such as quantum critical point at $x_{\rm cr}$.

In summary, we have narrowed down possible pair symmetry realized in (U,Th)Be$_{13}$.
The double transition is interpreted in terms of the 2D degeneracy scenario in E$_{\rm u}$.
The low and high  $T$ phases are identified as the cyclic $p$ wave state 
$\vec {d}({\bf k})={\hat x}k_x+\varepsilon{\hat y}k_y+\varepsilon^2{\hat z}k_z$ and the biaxial nematic state
${\vec d}({\bf k})=\sqrt3({\hat x}k_x-{\hat y}k_y)$ respectively. This leads us to simultaneously identify 
the uniaxial nematic state ${\vec d}({\bf k})=2{\hat z}k_z-{\hat x}k_x-{\hat y}k_y$ for undoped UBe$_{13}$.
Both orbital or gap structure and spin structure of the pairing functions for A, B, and C phases are
fully consistent with the recent~\cite{shimizu1} and the $\mu$SR~\cite{heffner,sonier} experiments.
We also put forth the accidental scenario previously proposed~\cite{sigrist} further 
and evaluated both scenarios critically.

Acknowledgments:
The author thanks Y. Shimizu, T. Sakakibara, and S. Kittaka for collaboration on
this project, which ultimately motivates this theory and Tom Rosenbaum, Y. Machida, and T. Mizushima for stimulating discussions.
K. Suzuki and H. Ikeda are acknowledged with their encouragement.
This work is supported by KAKENHI No.26400360 and No.17K05553 from JSPS
and also performed in part at the Aspen Center for Physics, 
which is supported by National Science Foundation grant PHY-1066293.


\begin{thebibliography}{99}

\bibitem{review1}
C. Pfleiderer,   
Rev. Mod. Phys. {\bf 81}, 1551 (2009).

\bibitem{review2}
T. Mizushima, Y. Tsutsumi, T. Kazwakami, M. Sato, M. Ichioka, and K. Machida,
J. Phys. Soc. Jpn. {\bf 85}, 022001  (2016).


\bibitem{tsutsumi}
See, for example, 
Y. Tsutsumi, M. Ishikawa, T. Kawakami, T. Mizushima, M. Sato, M. Ichioka, and K. Machida,
J. Phys. Soc. Jpn. {\bf 82}, 113707  (2013).

\bibitem{machida}
Y. Machida, A. Itoh, Y. So, K. Izawa, Y. Haga, E. Yamamoto, N. Kimura, Y. Onuki, Y. Tsutsumi, and K. Machida,
Phys. Rev. Lett. {\bf 108}, 157002 (2012).

\bibitem{sauls0}
J. A. Sauls,
Adv. Phys.  {\bf 43}, 113 (1994).

\bibitem{nishira}
K. Machida, T. Nishira, and T. Ohmi,
J. Phys. Soc. Jpn. {\bf 68}, 3364 (1999).

 \bibitem{sigrist-ueda}
 See review and references therein: M. Sigrist and K. Ueda, 
 Rev. Mod. Phys. {\bf 63}, 239 (1991). 

\bibitem{ott}
H. Ott, H. Rudigier, Z. Fisk, and J. L. Smith, 
Phys. Rev. B {\bf 31}, 1651 (1985).

\bibitem{batlogg}
B. Batlogg, D. Bishop, B. Golding, C. M. Varma, Z. Fisk, J. L. Smith, and H. R. Ott, 
Phys. Rev. Lett. {\bf 55},  1319 (1985).

\bibitem{heffner}
J. E. Sonier, R. H. Heffner, D. E. MacLaughlin, G. J. Nieuwenhuys, O. Bernal, R. Movshovich, P. G. Pagliuso, 
J. Cooley, J. L. Smith, and J. D. Thompson,
Phys. Rev. Lett.  {\bf 85}, 2821 (2000). 

\bibitem{sonier}
J. E. Sonier, R. H. Heffner, G. D. Morris, D. E. MacLaughlin, O. O. Bernal, J. Cooley, J. L. Smith, and J. D. Thompson,
Physica B  {\bf 326}, 414 (2003). 

\bibitem{shimizu1}
Y. Shimizu, S. Kittaka, S. Nakamura, T. Sakakibara, D. Aoki, Y. Homma, A. Nakamura, and K. Machida, 
Phys. Rev. B {\bf 96}, 100505(R)  (2017).


\bibitem{shimizu}
Y. Shimizu, S. Kittaka, T. Sakakibara, Y. Haga, E. Yamamoto, H. Amitsuka, Y. Tsutsumi
and K. Machida, Phys. Rev. Lett. {\bf 114}, 147002 (2015).


\bibitem{kromer}
F. Kromer, M. Lang, N. Oeschler, P. Hinze, C. Langhammer, F. Steglich, J. S. Kim, and G. R. Stewart,
Phys. Rev. B {\bf 62}, 12477 (2000).


 \bibitem{sauls}
J. A. Sauls,   J. Low Temp. Phys. {\bf 95}, 153 (1994).


\bibitem{sigrist}
M. Sigrist and T. M. Rice,
Phys. Rev. B {\bf 39}, 2200 (1989).


\bibitem{heffner0}
R. H. Heffner, J. L. Smith, J. O.  Willis, P. Birrer, C. Baines, F. N. Gygax, B. Hitti, E. Lippelt, H. R. Ott, 
A. Schenck, E. A. Knetsch, J. A. Mydosh, and D. E. MacLaughlin,
Phys. Rev. Lett. {\bf 65}, 2816 (1990).

\bibitem{kato1}
K. Machida and M. Kato,
Phys. Rev. Lett. {\bf 58}, 1986 (1987).


\bibitem{kato2}
M. Kato and K. Machida,
J. Phys. Soc. Jpn. {\bf 56}, 2136 (1987).


\bibitem{kato3}
M. Kato and K. Machida,
Phys. Rev. B {\bf 37}, 1510 (1988).

\bibitem{magne}
A. Hiess, R. H. Heffner, J. E. Sonier, G. H. Lander, J. L. Smith, and J. C. Cooley,
Phys. Rev. B {\bf 66}, 064531 (2002).


\bibitem{leggett}
A. J. Leggett, Rev. Mod. Phys. {\bf 47}, 331 (1975).
   
\bibitem{zieve}
R. J. Zieve, D. S.  Jin, T. F. Rosenbaum, J. S. Kim, and G. R. Stewart,
Phys. Rev. Lett. {\bf 72}, 756 (1994). 
 
 \bibitem{ozaki}
 M. Ozaki, K. Machida, and T. Ohmi, 
 Prog. Theor. Phys. {\bf 74}, 221 (1985). 

 
 \bibitem{gorkov}
 G. E. Volovik and L. P. Gor'kov,
 Sov. Phys. JETP {\bf 61}, 843 (1985). 
[Zh. Eksp. Teor. Fiz. {\bf 88}, 1412 (1985)].
 
 
 \bibitem{ueda}
 K. Ueda and T. M. Rice,
 Phys. Rev. B {\bf 31}, 7114 (1985).


 
\bibitem{take}
 K. Takegahara and H. Harima,
  Physica B  {\bf 281}-{\bf 282}, 764 (2000).
 
 \bibitem{maehira}
 T. Maehira, A. Higashiya, M. Higuchi, H. Yasuhara, and A. Hasegawa, 
 Physica B  {\bf 312}-{\bf 313}, 103 (2002).

 \bibitem{tamagaki}
R. Tamagaki, Prog. Theor. Phys.  {\bf 44}, 905 (1970).

\bibitem{mizushima1}
T. Mizushima, K. Masuda, and M. Nitta,
 Phys. Rev. B {\bf 95}, 140503 (R) (2017)
 and references therein.

\bibitem{furusaki}
A. P. Schnyder, S.Ryu, A. Furusaki, and A. W. W. Ludwig, 
 Phys. Rev. B {\bf 78}, 195125 (2008).
 
 \bibitem{mizushima2}
T. Mizushima and M. Nitta, arXiv:1710.07403.

 \bibitem{matano}
 K. Matano, M. Kriener, K. Segawa, Y. Ando, snf G. -Q. Zheng, 
 Nature Phys. {\bf 12}, 852 (2016).
 
  \bibitem{yonezawa}
  S. Yonezawa, K. Tajiri, S. Nakata, Y. Nagai, Z. Wang, K. Segawa, Y. Ando, and Y. Maeno,
  Nature Phys. (2016) doi 10.1038/nphys3907.
  
   \bibitem{visser}
   A. M. Nikitin, Y. Pan, Y. K. Huang, T. Naka, and A. de Visser,
  Phys. Rev. B {\bf 94}, 144516 (2016).
  
   
 \bibitem{fu}
 J. W. F. Venderbos, V. Kozii, and L. Fu, 
 Phys. Rev. B {\bf 94}, 094522 (2016).
 Y. Wang and L. Fu, arXiv:1703.06880.


  \bibitem{machida0}
 K. Machida, T. Ohmi, and M. Ozaki,
 J. Phys. Soc. Jpn. {\bf 54}, 1552 (1985).
 
 \bibitem{rosenbaum}
 Note that for $x=2.2\%$ sample used for the identical poly-crystalline sample as in Kromer~\cite{kromer},
 Jin et al suggested a crossing. But later Kromer's refined experiment exhibits no crossing.
 D. S. Jin, S. A. Carter, T. F. Rosenbaum, J. S. Kim, and G. R. Stewart, 
 Phys. Rev. B {\bf 53}, 8549 (1996).
The author thanks T. Rosenbaum for pointing it out.
   
\bibitem{machida2} 
  K. Machida, M. Ozaki, and T. Ohmi,
 J. Phys. Soc. Jpn. {\bf 58}, 4116 (1989). 
 
 \bibitem{machida3} 
K. Machida and M. Ozaki,
 Phys. Rev. Lett. {\bf 66}, 3293 (1991).
 
 
 
 
 
 
 
 
 
    
\end{thebibliography}

\end{document}